\documentclass{article}
\usepackage{amstext}
\usepackage{graphicx}

\newcommand{\beq}{\begin{eqnarray}}
\newcommand{\eeq}{\end{eqnarray}}

\newcommand{\dev}{\mathrm d}
\newcommand{\I}{\imath}
\newcommand{\E}{\mathrm{e}^}
\newcommand{\Tr}{\mathrm{Tr}}
\newcommand{\re}{\mathrm{Re}}

\newcommand{\im}{\mathrm{Im}}
\newcommand{\mS}{\mathcal{S}}
\newcommand{\mR}{\mathcal{R}}
\newcommand{\mP}{\mathcal{P}}
\newcommand{\mF}{\mathcal{F}}
\newcommand{\mG}{\mathcal{G}}

\newcommand{\Ci}{\textrm{C}i}
\newcommand{\si}{\textrm{s}i}

\newcommand{\hc}{\mathrm{h.c.}}

\newcommand{\Q}{\dot{\mathcal{Q}}}

\newcommand{\Og}{\mathcal{O}}

\begin{document}

\title{Atom-wall dispersive forces from master equation formalism}
\author{T.N.C. Mendes\footnote{tarciro@if.uff.br}$\;$ and C. Farina\footnote{farina@if.ufrj.br} 
\\ \textit{Instituto de F\'isica - UFF, Brazil}\\
\textit{Instituto de F\'isica - UFRJ, Brazil}
}
\date{December 19, 2006}         % Enter your date or \today between curly braces
\maketitle
\begin{abstract}

Using the general expressions for level shifts obtained from the {\it master equation} for a small system interacting with a large one considered as a reservoir, we calculate the dispersive potentials between an atom and a wall in the dipole approximation. We analyze in detail the particular case of a two-level atom in the presence of a perfectly conducting wall. We study the van der Waals as well as the re\-so\-nant interactions. All distance regimes as well as the high and low temperature regimes are considered. We show that the Casimir-Polder interaction can not be considered as a direct result of the vacuum fluctuations {\it only}. Concerning the interaction between the atom and the wall at high temperature, which show that a saturation of the potential for all distances occurs. This saturated potential coincides exactly with that obtained in the London-van der Waals limit.
\end{abstract}
%
%
%%%%%%%%%%%%%%%%%%%%%%%%%%%%%%%%%%%%%%%%%%%%%%%%%%%%%%%%%%%%%%%%%%%%%%%%%%%%%%%%%%%%%%%%%%%%%%%%%%%%%%%%%
%%%%%%%%%%%%%%%%%%%%%%%%%%%%%%%%%%%%%%%%%%%%%%%%%%%%%%%%%%%%%%%%%%%%%%%%%%%%%%%%%%%%%%%%%%%%%%%%%%%%%%%%%
%
\section{Introduction}      % Enter section title between curly braces

The subject of dispersive interactions is an old problem related to the intermolecular forces whose discussion first eppeared in the van der Waals state equation for real gases in 1873 \cite{VDWT}. In this equation there is a term proportional to $1/v^2$, where $v$ is the specific volume of the gas, responsible for a long-range interaction between the molecules. This interaction provides a mechanism for the phase transitions between the gas and liquid states (and {\it vice-versa}) of the substance. However, only in the beginning of the 20th century, these interactions were understood \cite{Margenau}.

Approximately a hundred years ago, M. Reinganum \cite{Reig1903} suggested that van der Waals forces could be explained by the electromagnetic interaction between the electric dipoles
associated to the molecules. In the twenties, mainly due to W.H. Keesom and P. Debye works \cite{Keesom,Debye}, Reinganum's insight showed to be very fruitful, making possible the explanation of thevan der Waals interaction between two polar molecules ({\it orientation forces}) and between a polar molecule and a nonpolar one ({\it induced forces}). 

However, the explanation for the force between neutral and non-polar mole\-cu\-les, like noble gases for example, had to wait for the advent of Quantum Mechanics. In 1930, F. London \cite{London} obtained the interaction potential between two neutral hydrogen atoms in their respective ground states  and separated by a distance much larger than Bohr radius in the static electric dipole limit, $\lambda_0\gg r\gg a_0$, where $\lambda_0 = 2\pi c/\omega_0$ is the wavelength of the main atomic  transition relevant to the interaction  and $a_0$ is the Bohr radius. London's result is 
$V(r)\simeq -6.5e^2 a_0^5/r^6\simeq -3\hbar\omega_0\alpha_0^2/4r^6$.

As one can see, this interaction is a genuine quantum effect (proportional to $\hbar$) %
and depends only on the atomic polarizabilities. Since the polarizability is intimately connected to %
dispersive part of the electric permittivity of the substance, this kind of interaction is called %
{\it dispersive van der Waals interaction} \cite{Langbein}. Then, dispersive van der Waals forces are long-range electromagnetic forces between non-polar and neutral molecules in their ground states.

In 1948, Casimir and Polder \cite{CasPol1948} considered for the first time the influence of retardation effects on the van der Waals forces between two atoms as well as on the force between an atom and a perfectly conducting wall. They showed that in the retarded limit the interaction falls as $1/r^7$ for two atoms (in contrast to London's $1/r^6$ result) and as $1/r^4$ for an atom and a perfectly conducting wall, in contrast with the short distance limit (that falls as $1/r^3$) \cite{Len}.

In 1956, E.M. Lifshitz developed a general theory of dispersive van der Waals forces between dielectric macroscopic bodies using a non-perturbative approach \cite{Lifshitz}, because the perturbative expansion for many-body interaction breaks down due to the non-additivity of the van der Waals forces. Lifshitz derived a powerful expression for the force at finite temperature between two semi-infinite dispersive media characterized by an electric permittivity and separated by a slab of any other dispersive medium. He was able to derive and predict several results, like the variation of the thickness of thin superfluid helium films in a remarkable agreement with experiments \cite{SabAnd}. He also showed that the Casimir-Polder force is a limiting case when one of the media is sufficiently dilute such that the force between the slabs may be obtained by direct  pair-wise integration of a single atom-wall interaction \cite{DLP}.

Since then a wide knowledge about the behavior amd nature of dispersive forces has been achieved. In the following decades, many works have been done about this subject, like those done by A.D. McLachlan \cite{McLT=0}, in the 60's, where the recently compiled linear response theory developed by R. Kubo \cite{Kubo1954} was applied. In 1970, G. Feinberg and J. Sucher \cite{FeinSucher} treated the electric and magnetic contributions to van der Waals interaction in the same foot. In the last 60's decade and during the 70's it was analyzed the interaction between excited atom and a wall \cite{Morawitz,Barton72,Barton74}. The first of these works, made by H. Morawitz, considered the level and frequency shifts of an excited atom in the presence of a perfectly conducting wall using the image method, which led him to the discovery of the {\it resonant dispersive interaction}, that we will discuss later. In the 80's and 90's the non-additivity character of the van der Waals forces was exhaustively exploited \cite{Thiru,Thiru94}; very good works concerning level shifts of atoms and dispersive interactions in cavities \cite{Meschede,Hinds,Jhe,Jhe2,Nha} and the thermal contributions to the Casimir-Polder interaction between atoms and finite systems (like cylinders or spheres), in a microscopic approach, was studied \cite{GoeRoy,BartCPFS,BartS}.

More recently the interest for van der Waals and Casimir-Polder forces have been raised mainly due to better experimental techniques and the development of nanosciences. The influence of these forces in the stability of Bose-Einstein condensates \cite{Vuletic,Antezza} and on factoring and doping carbon nanotubes are branches of great activity nowadays \cite{Nan1,Nan2,Nan3}.

In this paper we make the sequence of the work began in a previous one \cite{TarFarJPA2006}, where we studied the van der Waals interaction between a two-level atom and a perfectly conducting wall using the density matrix formalism \cite{DDC1984}. Here, we come back to this problem using the same formalism, but studying also the excited state contributions to the interaction, both in vacuum and thermal states of the electromagnetic field. We give simple interpretations to the ground state contribution (van der Waals interaction) and the excited state contribution (resonant interaction) in terms of the non-resonant and resonant parts of the atomic polarizability. Considering the thermal state of the electromagnetic field, we analyze both low and hight temperature limits, $\hbar\omega_0\ll k_B T$ and $\hbar\omega_0\gg k_B T$ respectively, and show that in the last limit the interaction between the two-level system and the perfectly conducting wall is temperature independent and behaves exactly like that in the London-van der Waals limit for vacuum contribution.
\section{Level shifts and exchange energy rates}
In this section we obtain the level shifts and the exchange energy rates of a multi-level system interacting with the electromagnetic field in the dipole approximation. In the first case, if we  consider the boundary conditions imposed to the field by the presence of a given wall, we can extract the position dependent part, which gives, by derivation, the atom-wall dispersive forces.

\subsection{General expressions}

Our starting point is the master equation for a small system $\mS$ (the atom) which interacts weakly with a large one $\mR$, that may be considered as a reservoir (quantized electromagnetic field), by the interaction hamiltonian
\begin{equation}
\label{Vint}
V = -\sum_j S_j R_j\,,
\end{equation}
where $S_j$ and $R_j$ are observables associated to $\mS$ and $\mR$ respectively. In the Schr\"odinger picture, the master equation may be written as \cite{DDC1984,DDC1982,AtPhInt}
\beq
\label{eqmestra}
{\dev\over \dev t}\rho_{ab}^S\left(t\right) &=& -\I\omega_{ab}\rho_{ab}^S\left(t\right) + \sum_{c,d}\mathcal J_{abcd}\rho_{cd}^S\left(t\right)\, ,
\\
%%%%%%%%%%%%%%%%%%%%%%%%%%%%%%%%%%%%%%%%%%%%%%%%%%%%%%%%%%%%%%%%%%%%%%%%%%%%%%
%
\label{txmestra}
{\mathcal J}_{abcd}&=&-\frac{1}{\hbar^2}\sum_{j,k}\int_0^\infty \dev\tau\Bigg\lbrace g_{jk}^R\left( \tau\right) \left[ \delta_{bd}\sum_n S_{an}^j S_{nc}^k \E{-\I\omega_{nc}\tau} - S_{ac}^k S_{db}^j \E{-\I\omega_{ac}\tau}\right]+
\cr
&+& g_{kj}^R\left( -\tau\right) \left[ \delta_{ac} \sum_n S_{dn}^k S_{nb}^j \E{\I\omega_{nd}\tau} - S_{ac}^j S_{db}^k \E{\I\omega_{bd}\tau}\right]\Bigg\rbrace\,,
\eeq
where $\rho_{ab}^S = \langle a\vert\rho_S\vert b\rangle$ is the matrix element of the density operator $\rho_S$, associated to the system $\mS$, between the energy eigenstates $\vert a\rangle$ and $\vert b\rangle$ with eigenvalues $E_a$ and $E_b$ of the unperturbed hamiltonian operator $H_S$ of $\mS$ ($H_S\vert a\rangle = E_a\vert a\rangle$); $S_{ab}^j = \langle a\vert S_j\vert b\rangle$ is the matrix element of the observable $S_j$ and $\omega_{ab} = \left(E_a-E_b\right)/\hbar$ is the transition frequency between the states $\vert a\rangle$ and $\vert b\rangle$. The function $g_{jk}^R\left( \tau\right)$ that appears in equation (\ref{txmestra}) is defined as
\beq
\label{gtau_jk}
g_{jk}^R\left(\tau\right) = \left[g_{kj}^R\left(-\tau\right)\right]^{*} = \Tr_R\left[\rho_R R_j\left(\tau\right) R_k\left(0\right)\right] =  \sum_{\mu}p_{\mu}\sum_{\nu}R_{\mu\nu}^j R_{\nu\mu}^k \E{\I\omega_{\mu\nu}\tau}\, ,
\eeq
where $R_{\mu\nu}^j = \langle\mu\vert R_j\vert\nu\rangle$ is the matrix element of the observable $R_j$ between the energy eigenstates $\vert\mu\rangle$ and $\vert\nu\rangle$ with energy eigenvalues $E_{\mu}$ and $E_{\nu}$ of the unperturbed hamiltonian operator $H_R$ of $\mR$ ($H_R\vert\mu\rangle = E_{\mu}\vert\mu\rangle$); $\omega_{\mu\nu} = \left(E_{\mu}-E_{\nu}\right)/\hbar$ is the transition frequency between the states $\vert\mu\rangle$ and $\vert\nu\rangle$ and $\rho_R$ is the density matrix associated to $\mR$, considered constant in time and diagonal in the $\{\vert\mu\rangle\}$ base-ket, so that
\beq
\label{pR}
\rho_R = \sum_{\mu}p_{\mu}\vert\mu\rangle\langle\mu\vert\,,
\eeq
where $p_{\mu}$ is the statistical weight of the state $\vert\mu\rangle$ for a given {\it ensemble}.

Making a brief resume, equations (\ref{eqmestra}-\ref{txmestra}) were obtained from the equation of the time evolution of the total density matrix $\rho\left(t\right)\simeq \rho_S\left(t\right)\otimes\rho_R$, $${\dev\over\dev t}\rho\left(t\right) = {\I\over\hbar}\left[\rho\left(t\right),H\right]\,,$$ where $H = H_S+H_R+V$ is the hamiltonian of the total system $\mS+\mR$. Then, one has made implicitly the assumption that there are two very different time scales in the evolution of the total system $\mS+\mR$: the characteristic time $T_S$  in which the average values of the observables of $\mS$ change significantly and the characteristic time $\tau_c$, which is, crudely speaking, the width of $g_{jk}^R\left(\tau\right)$ and measures the time of the fluctuations of the reservoir observables.

The approximations used in derivation of the equations refered above are based on the following conditions,
\beq
\label{Markov}
\tau_c\ll \Delta t \ll T_S\, ,
\eeq
where $\Delta t$ is the time interval that enters in the calculation of the time derivative in equation (\ref{eqmestra}). Last condition tells us that the master equation is a {\it coarse grained} rate, since rapid variations of $\rho_S$ that occur in times of order of $\tau_c$ are smoothed in the interval $\Delta t$; since we are interest in times of the order of $T_S$, this smoothing remains a good aproximation once condition (\ref{Markov}) is satisfied. Another suplementary condition for obtaining equation (\ref{eqmestra}), which makes the coefficients $\mathcal I_{abcd}$ independent of $\Delta t$, is that frequency difference between the elements of the density matrix should be very small compared to $1/\Delta t$, so that $$\vert\omega_{ac}-\omega_{bd}\vert\Delta t\ll 1\,.$$
Last condition is called {\it secular condition} and the sum in equation (\ref{eqmestra}) is only over secular terms (terms that satisfy the secular condition).

Using two new assumptions, one may conveniently split equation (\ref{eqmestra}) into two other ones: an equation for the diagonal elements $\rho^S_{aa}$ (populations) and another for the non-diagonal elements $\rho^S_{ab}$, with $a\not= b$ (coherences). For the populations, the assumption is that there is no coherence $\rho^S_{cd}$ in the sum in equation (\ref{eqmestra}) with too low frequency, so that $\omega_{cd} T_S\gg 1$. This implies that all coherences are not secular terms and the master equation for the populations $\rho^S_{aa}$ may be put into the form
\beq
\label{EQMP}
{\dev\over \dev t}\rho_{aa}^S\left(t\right) &=& \sum_{c}\left(\rho_{cc}^S\left(t\right)\Gamma_{c\rightarrow a}-\rho_{aa}^S\left(t\right)\Gamma_{a\rightarrow c}\right)\,,
\\
%%%%%%%%%%%%%%%%%%%%%%%%%%%%%%%%%%%%%%%%%%%%%%%%%%%%%%%%%%%%%%%%%%%%%%%%%%%%%%%%%%%%%%%%%%
%
\label{Fermi}
\Gamma_{c\rightarrow a}&=&{2\pi\over\hbar}\sum_{\mu}p_{\mu}\sum_{\nu}\vert\langle \mu,c\vert V\vert \nu,a\rangle\vert^2\delta\left(E_{\mu}+E_c-E_{\nu}-E_{a}\right)\,,\;\;\;
\eeq
where $V$ is the interaction hamiltonian given by (\ref{Vint}). The quantity $\Gamma_{c\rightarrow a}$ may be interpreted as the transition rate probability between the states $\vert c\rangle$ and $\vert a\rangle$ as a result of the interaction of the system with the reservoir.

In order to obtain the equation for coherences we make the second assumption, namely, we consider only the non-degenerate case: the frequency $\omega_{ab}$ differ from all other frequencies $\omega_{cd}$ by a quantity of order (or larger than) $1/\Delta t$, that is, $\vert\omega_{ac}-\omega_{bd}\vert\sim 1/\Delta t\gg 1/T_S$. Then, the only secular term in equation (\ref{eqmestra}) is that which couples the coherence $\rho^S_{ab}$ with itself, so that the equation for this coherence may be written as
\beq
\label{eqcoe}
{\dev\over \dev t}\rho_{ab}^S\left(t\right) = -\I\left(\omega_{ab}+\Delta_{ab}\right)\rho_{ab}^S\left(t\right) - \Gamma_{ab}\rho_{ab}^S\left(t\right)\;,\;\;\;\;\;\;a\not=b
\eeq
where
\beq
\label{Dab}
\Delta_{ab}&=& \Delta_{a}-\Delta_{b}
\\
%%%%%%%%%%%%%%%%%%%%%%%%%%%%%%%%%%%%%%%%%%%%%%%%%%%%%%%%%%%%%%%%%%%%%%%
%
\label{Gab}
\Gamma_{ab} &=& \Gamma_{ab}^{\mathrm{ad.}}+\Gamma_{a}^{\mathrm{nad.}}+\Gamma_{b}^{\mathrm{nad.}}
\\
%%%%%%%%%%%%%%%%%%%%%%%%%%%%%%%%%%%%%%%%%%%%%%%%%%%%%%%%%%%%%%%%%%%%%%%
%
\label{Dshift}
\Delta_{n}&=& {1\over\hbar^2}\sum_{\mu}p_{\mu}\sum_{\nu}\sum_{j}\mathcal P{\vert\langle\mu,n\vert V\vert j,\nu\rangle\vert^2\over\omega_{\mu\nu}-\omega_{jn}}
\\
%%%%%%%%%%%%%%%%%%%%%%%%%%%%%%%%%%%%%%%%%%%%%%%%%%%%%%%%%%%%%%%%%%%%%%%
%
\label{Gad}
\Gamma_{ab}^{\mathrm{ad.}}&=&-{2\pi\over\hbar^2}\sum_{\mu}p_{\mu}\sum_{\nu}\langle\mu,b\vert V\vert b,\nu\rangle\langle\nu,a\vert V\vert a,\mu\rangle\delta\left(\omega_{\mu\nu}\right)
\\
%%%%%%%%%%%%%%%%%%%%%%%%%%%%%%%%%%%%%%%%%%%%%%%%%%%%%%%%%%%%%%%%%%%%%%%
%
\label{Gnad}
\Gamma_{n}^{\mathrm{nad.}}&=& {\pi\over\hbar^2}\sum_{\mu}p_{\mu}\sum_{\nu}\sum_{j}\vert\langle\mu,n\vert V\vert j,\nu\rangle\vert^2\delta\left(\omega_{\mu\nu}-\omega_{jn}\right)={1\over 2}\sum_{j}\Gamma_{n\rightarrow j}\,\;\;\;\;\;\;\;\;
\eeq
and the symbol $\mP$ in equation (\ref{Dshift}) represents the Cauchy principal value.

The integration of the equation (\ref{eqcoe}) is a straightforward task and gives $$\rho_{ab}^S\left(t\right) = \rho_{ab}^S\left(0\right)\E{-\Gamma_{ab}\, t}\E{-\I\left(\omega_{ab}+\Delta_{ab}\right)\,t}\,,$$ from which one may conclude that the coherence $\rho^S_{ab}$, as a result of its interaction with the reservoir, oscillate with a frequency shifted by $\Delta_{ab}$ from its free value $\omega_{ab}$ and decays exponentially with a characteristic time given by $1/\Gamma_{ab}$. From equations (\ref{EQMP}-\ref{Fermi}) and (\ref{Dshift}) one may obtain the exchange energy rates between the system and the reservoir and the energy level shifts of the system, respectively, since frequency shifts result from levels shifts. Recall that $\delta E_a = \hbar\Delta_a$ is the shift in the energy of the state $\vert a\rangle$ (and equivalently for $\vert b\rangle$).

For energy rates, considering the time derivative of the average value of the hamiltonian of the system in Schr\"odinger and Heisenberg pictures, one will find
\beq
\label{txHa}
{\dev\langle H_S\rangle_a\over \dev t} = \sum_{b}\left(E_b-E_a\right)\Gamma_{a\rightarrow b}=\sum_{b}\hbar\omega_{ba}\Gamma_{a\rightarrow b}\,,
\eeq
which may be interpreted as a net rate of changing the average value of the energy of the system $\mS$ when it is in the state $\vert a\rangle$ or, in other words, the {\it exchange energy rate} between $\mS$ and $\mR$ for the state $\vert a\rangle$. It is possible to put last result into a more intuitive form that clearly exhibits the roles played by $\mS$ and $\mR$. Returning to equation (\ref{gtau_jk}), one may show that real and imaginary parts of $g_{jk}^R\left(\tau\right)$ are related to symmetric correlation function $C_{jk}^R\left(\tau\right)$ and the linear susceptibility $\chi_{jk}^R\left(\tau\right)$ \cite{AtPhInt,Kubo66}
\beq
\label{Ctau}
C_{jk}^R\left(\tau\right)\! &=&\! \re\left[g_{jk}^R\left(\tau\right)\right] = \sum_{\mu}p_{\mu}\sum_{\nu} R_{\mu\nu}^j R_{\nu\mu}^k\cos\left(\omega_{\mu\nu}\tau\right)\,,
\\
%%%%%%%%%%%%%%%%%%%%%%%%%%%%%%%%%%%%%%%%%%%%%%%%%%%%%%%%%%%%%%%%%%%%%%%%%%%%%
%
\label{Chi_tau}
\chi_{jk}^R\left(\tau\right)\! &=&\! {2\over\hbar}\Theta\left(\tau\right)\im\left[g_{jk}^R\left(\tau\right)\right] = -{2\over\hbar}\sum_{\mu}p_{\mu}\sum_{\nu}R_{\mu\nu}^{j} R_{\nu\mu}^{k}\Theta\left(\tau\right)\sin\left(\omega_{\mu\nu}\tau\right),\;\;\;\;\;\;
\eeq
where $\Theta\left(\tau\right)$ is the step function. In the frequency space, we have
\beq
\label{FC}
\hat C_{jk}^R\left(\omega\right) &=& \int_{-\infty}^{\infty}\dev\tau C_{jk}^R\left(\tau\right)\E{\I\omega\tau}
%
%\cr
%%%%%%%%%%%%%%%%%%%%%%%%%%%%%%%%%%%%%%%%%%%%%%%%%%%%%%%%%%%%%%%%%%%%%%%%%%
%
%\label{CRw}
%
= \pi \sum_{\mu}p_{\mu}\sum_{\nu}R_{\mu\nu}^{j} R_{\nu\mu}^{k}\Big[ \delta\left( \omega+\omega_{\mu\nu}\right) +\delta\left( \omega-\omega_{\mu\nu}\right) \Big]\; ,
\\
%%%%%%%%%%%%%%%%%%%%%%%%%%%%%%%%%%%%%%%%%%%%%%%%%%%%%%%%%%%%%%%%%%%%%%%%%%%%%%%%
%
%
\hat\chi_{jk}^R\left(\omega\right) &=& \int_{-\infty}^{\infty}\dev\tau \chi_{jk}^R\left(\tau\right)\E{\I\omega\tau} =
\hat\chi_{jk}^{\,\prime R}\left(\omega\right) + \I\hat\chi_{jk}^{\,\prime\prime R}\left(\omega\right)\, ,
\\
%%%%%%%%%%%%%%%%%%%%%%%%%%%%%%%%%%%%%%%%%%%%%%%%%%%%%%%%%%%%%%%%%%%%%%%%%%%
%
\label{Chi'R}
\hat\chi_{jk}^{\,\prime R}\left(\omega\right) &=& -\frac{1}{\hbar}\sum_{\mu}p_{\mu}\sum_{\nu}R_{\mu\nu}^{j} R_{\nu\mu}^{k}\left[ {\mathcal P}\frac{1}{\omega_{\mu\nu}+\omega}+{\mathcal P}\frac{1}{\omega_{\mu\nu}-\omega}\right]\, ,
\\
%%%%%%%%%%%%%%%%%%%%%%%%%%%%%%%%%%%%%%%%%%%%%%%%%%%%%%%%%%%%%%%%%%%%%%%%%%%
%
\label{Chi''R}
\hat\chi_{jk}^{\,\prime\prime R}\left(\omega\right) &=& \frac{\pi}{\hbar}\sum_{\mu}p_{\mu}\sum_{\nu}R_{\mu\nu}^{j} R_{\nu\mu}^{k}\Big[ \delta\left(\omega_{\mu\nu}+\omega\right)-\delta\left(\omega_{\mu\nu}-\omega\right)\Big]\;.
\eeq
The symmetric correlation function given by (\ref{FC}) may be interpreted as a dispersion of the observables $R_j$ and $R_k$ arround a frequency $\omega$. It is a measure of the {\it fluctuations} of the reservoir dynamical variables. The real part of the susceptibility defined in equation (\ref{Chi'R}) is called the {\it dispersive} or reactive part and is related to the polarization of a system by an external perturbation. On the other hand, the imaginary part is directly responsible for the absorption and dissipation of energy by the system and, for this reason is called {\it dissipative} part of susceptibility \cite{Landau}.

Now, let us come back to the equations (\ref{Dshift}) and (\ref{txHa}). Using the quantities defined in equations (\ref{FC}-\ref{Chi''R}), the energy level shifts and the exchange energy rates may be cast into the form

\beq
\label{dEa}
\delta E_a &=& \delta E_a^{fr} + \delta E_a^{rr}\, ,
\\
%%%%%%%%%%%%%%%%%%%%%%%%%%%%%%%%%%%%%%%%%%%%%%%%%%%%%%%%%%%%%%%%
%
\label{Qa}
{\dev\langle H_S\rangle_a\over \dev t} &=& \Q_a\; =\; \dot{\mathcal Q}_a^{fr} + \dot{\mathcal Q}_a^{rr}\,,
\eeq
\vskip -0.5 cm
\beq
\label{fr}
\delta E_a^{fr}&=&-\frac{1}{2}\sum_{j,k}\int_{-\infty}^{\infty}\frac{\dev\omega}{2\pi}\,\hat\chi_{jk}^{\,\prime S,a}\left(\omega\right) \hat C_{kj}^{R}\left( \omega\right)\,,
\\
%%%%%%%%%%%%%%%%%%%%%%%%%%%%%%%%%%%%%%%%%%%%%%%%%%%%%%%%%%%%%%%%%%%%%%%%%%
%
\label{rr}
\delta E_a^{rr}&=&-\frac{1}{2}\sum_{j,k}\int_{-\infty}^{\infty}\frac{\dev\omega}{2\pi}\,\hat\chi_{jk}^{\,\prime R}\left(\omega\right) \hat C_{kj}^{S,a}\left( \omega\right)\,,
\\
%%%%%%%%%%%%%%%%%%%%%%%%%%%%%%%%%%%%%%%%%%%%%%%%%%%%%%%%%%%%%%%%%%%%%%%%%%%
%
\label{dfr}
\dot{\mathcal Q}_a^{fr} &=& \sum_{j,k}\int_{-\infty}^{\infty}{\dev\omega\over 2\pi}\,\omega\,\hat\chi_{jk}^{\,\prime\prime\, S,a}\left(\omega\right)\hat C_{kj}^{R}\left(\omega\right)\,,
\\
%%%%%%%%%%%%%%%%%%%%%%%%%%%%%%%%%%%%%%%%%%%%%%%%%%%%%%%%%%%%%%%%%%%%%%
%
\label{drr}
\dot{\mathcal Q}_a^{rr} &=& -\sum_{j,k}\int_{-\infty}^{\infty}{\dev\omega\over 2\pi}\,\omega\,\hat\chi_{jk}^{\,\prime\prime\,R}\left(\omega\right)\hat C_{kj}^{S,a}\left(\omega\right)\,,
\eeq
where
\beq
\label{CSw}
\hat C_{jk}^{S,a}\left(\omega\right) &=& \pi \sum_{n} S_{a n}^{j} S_{n a}^{k}\Big[ \delta\left( \omega+\omega_{an}\right) +\delta\left( \omega-\omega_{an}\right) \Big]\,,
\\
%%%%%%%%%%%%%%%%%%%%%%%%%%%%%%%%%%%%%%%%%%%%%%%%%%%%%%%%%%%%%%%%%%%%%
%
\label{Chi'S}
\hat \chi_{jk}^{\,\prime S,a}\left(\omega\right)&=& -\frac{1}{\hbar}\sum_{n} S_{a n}^{j} S_{n a}^{k}\left[ {\mathcal P}\frac{1}{\omega_{a n}+\omega}+{\mathcal P}\frac{1}{\omega_{a n}-\omega}\right]\,,
\\
%%%%%%%%%%%%%%%%%%%%%%%%%%%%%%%%%%%%%%%%%%%%%%%%%%%%%%%%%%%%%%%%%%%%%%%%%%%%%%%%%%%%%
%
\label{Chi''S}
\hat\chi_{jk}^{\,\prime\prime\, S,a}\left(\omega\right) &=& \frac{\pi}{\hbar}\sum_{n}S_{an}^{j} S_{na}^{k}\Big[ \delta\left(\omega_{an}+\omega\right)-\delta\left(\omega_{an}-\omega\right)\Big]\,,
\eeq
are the symmetric correlation function and the dispersive and dissipative part of the susceptibility of the system $\mS$ in the state $\vert a \rangle$ respectively. Equations (\ref{fr}-\ref{drr}) are the general expressions for the level shifts and the exchange energy rates of a small system interacting with a reservoir.

The physical interpretation of equations (\ref{fr}-\ref{rr}) is simple. The former gives the contribution to the level shift due to polarization of the system $\mS$ by {\it fluctuations of the reservoir} $(fr)$ and the latter gives the contribution of the polarization of the reservoir by fluctuations of the system, or the {\it reservoir reaction} $(rr)$ contribution. This last case is analogous to the {\it radiation reaction} of an accelerated charge in an external field.

A similar interpretation may be done for equations (\ref{dfr}-\ref{drr}). The first of them may be understood as the {\it power absorbed} by the system $\mS$ from the reservoir fluctuations and the second represents the {\it power dissipated} by the system or, equivalently, the power lost to the reservoir.

\subsection{Dipole interacting with the radiation field}

Let us consider a small neutral but polarizable system which interacts weakly with the radiation field. Assuming the dipole approximation and considering the total quantized electric field operator at the positon $\mathbf{x}$ of the center of mass of the system as a sum over all possible modes,  the interacting hamiltonian takes the form
\beq
\label{Vint_mode}
V = -\mathbf{d}\cdot \mathbf{E}\left(\mathbf{x},t\right) = e\sum_{\mathbf{k}\lambda}\sum_j \left(x_j f_{\mathbf{k}\lambda}^j\left(\mathbf{x}\right)\E{\I\omega_k t} a_{\mathbf{k}\lambda}^{\dag}+\hc\right)\;,
\eeq
where $\mathbf{d} = -e\mathbf{r}$ is the dipole moment operator of the system, $-e$ is the electrical charge, $\mathbf{r} = \left(x_1,x_2,x_3\right)$ is the position operator of the charge and the functions $\mathbf{f}_{\mathbf{k}\lambda}\left(\mathbf{x}\right) = \hat x_j f_{\mathbf{k}\lambda}^j\left(\mathbf{x}\right)$ (we are using Einstein's convention here), where $\hat x_j$ is the unitary vector in the direction of component $x_j$ of $\mathbf{r}$, carry the information about the boundary conditions and possible source contributions. The index $j$ runs over cartesian components $\left(j = 1,2,3\right)$ and the index $\mathbf{k}\lambda$ specifies a field mode, where $\mathbf{k}$ is its wave-vector and $\lambda$ is the corresponding polarization. Operators $a_{\mathbf{k}\lambda}$ and $a_{\mathbf{k}\lambda}^{\dag}$ are the annihilation and creation operators, respectively, of one photon in the mode $\mathbf{k}\lambda$ and satisfies the usual commutation relations
\beq
\label{Com_aa}
\big[a_{\mathbf{k}\lambda},a_{\mathbf{k}^{\prime}\lambda^{\prime}}\big] = \big[a_{\mathbf{k}\lambda}^{\dag},a_{\mathbf{k}^{\prime}\lambda^{\prime}}^{\dag}\big] = 0\;,\;\;
%
%\\
%%%%%%%%%%%%%%%%%%%%%%%%%%%%%%%%%%%%%%%%%%%%%%%%%%%%%%%%%%%%%%%%%%
%\label{Com_aa+}
%
\big[a_{\mathbf{k}\lambda},a_{\mathbf{k}^{\prime}\lambda^{\prime}}^{\dag}\big] = \delta_{\mathbf{k}\mathbf{k}^{\prime}}\delta_{\lambda\lambda^{\prime}}\,.
\eeq

Choosing the coordinate axis in a manner such that the susceptibility and the correlation function of the system are diagonal we have
\beq
\label{CSdjk}
\hat C_{jk}^{S,a}\left(\omega\right)&=&\hat C_{aj}\left(\omega\right)\delta_{jk}\,,
\\
%%%%%%%%%%%%%%%%%%%%%%%%%%%%%%%%%%%%%%%%%%%%%%%%%%%%%%%%%%%%%%%%%%%%%%%%%%%%%%%%%%
%
\label{XS'djk}
\hat\chi_{jk}^{\,\prime S,a}\left(\omega\right) &=& \alpha_{aj}^{\,\prime}\left(\omega\right)\delta_{jk}\,,
\\
%%%%%%%%%%%%%%%%%%%%%%%%%%%%%%%%%%%%%%%%%%%%%%%%%%%%%%%%%%%%%%%5
%
\label{XS''djk}
\hat\chi_{jk}^{\,\prime\prime S,a}\left(\omega\right) &=& \alpha_{aj}^{\,\prime\prime}\left(\omega\right)\delta_{jk}\,,
\eeq
where
\beq
\hat C_{aj}\left(\omega\right) &=& \pi\hbar\sum_b{\alpha^{j}_{ab}\omega_{ba}\over 2}\Big[ \delta\left(\omega_{ba}+\omega\right)+\delta\left(\omega_{ba}-\omega\right)\Big]\,,
\\
%%%%%%%%%%%%%%%%%%%%%%%%%%%%%%%%%%%%%%%%%%%%%%%%%%%%%%%%%%%%%%%%%%%%%%%%%%%%%%%%%%
%
\label{a'}
\alpha_{aj}^{\,\prime}\left(\omega\right) &=& \sum_b{\alpha^{j}_{ab}\omega_{ba}\over 2}\left[ \mP {1\over \omega_{ba}+\omega}+\mP {1\over \omega_{ba}-\omega}\right]\,,
\\
%%%%%%%%%%%%%%%%%%%%%%%%%%%%%%%%%%%%%%%%%%%%%%%%%%%%%%%%%%%%%%%%%%%%%%%
%
\label{a''}
\alpha_{aj}^{\,\prime\prime}\left(\omega\right) &=& \pi\sum_b{\alpha^{j}_{ab}\omega_{ba}\over 2}\Big[ \delta\left(\omega_{ba}-\omega\right)-\delta\left(\omega_{ba}+\omega\right)\Big]\,,
\eeq
and
\beq
\label{a_static}
\alpha^{j}_{ab} = -{2e^2\over \hbar\omega_{ab}}\vert \langle a\vert x_j\vert b\rangle\vert^2
\eeq
is the static polarizability of the system in the direction $\hat x_j$ between the states $\vert a \rangle$ and $\vert b \rangle$.

Using last equations in (\ref{fr}-\ref{drr}), making $\vert\mu\rangle = \vert n_{\mathbf{k}\lambda}\rangle$ in equation (\ref{pR}) (which means a Fock state with $n$ photons in the mode $\mathbf{k}\lambda$) and performing the integration on $\omega$, we obtain, after some calculations, 
\beq
\label{dEa_rr}
\delta E^{rr}_a &=& -{1\over 2}\sum_j \sum_{\mathbf{k}\lambda}\alpha_{aj}^{\,\prime(-)}\left(k\right)\vert f^j_{\mathbf{k}\lambda}\left(\mathbf{x}\right)\vert^2 \,,
\\
%%%%%%%%%%%%%%%%%%%%%%%%%%%%%%%%%%%%%%%%%%%%%%%%%%%%%%%%%%%%%%%%%%%%%%%%
%
\label{Qa_rr}
\dot{\mathcal Q}_a^{rr} &=& - \sum_j \sum_{\mathbf{k}\lambda}c\,k\,\alpha_{aj}^{\,\prime\prime(-)}\left(k\right)\vert f^j_{\mathbf{k}\lambda}\left(\mathbf{x}\right)\vert^2 \,,
\\
%%%%%%%%%%%%%%%%%%%%%%%%%%%%%%%%%%%%%%%%%%%%%%%%%%%%%%%%%%%%%%%%%%%%%%%
%
\label{dEa_fr}
\delta E^{fr}_a &=& -\sum_j \sum_{\mathbf{k}\lambda} \alpha_{aj}^{\,\prime(+)}\left(k\right)\vert f^j_{\mathbf{k}\lambda}\left(\mathbf{x}\right)\vert^2 \left( \langle n_{\mathbf{k}\lambda}\rangle + {1\over 2}\right)\,,
\\
%%%%%%%%%%%%%%%%%%%%%%%%%%%%%%%%%%%%%%%%%%%%%%%%%%%%%%%%%%%%%%%%%%%%%%%%%%%%%%%%
%
\label{Qa_fr}
\dot{\mathcal Q}_a^{fr} &=& \sum_j \sum_{\mathbf{k}\lambda} c\,k\,\alpha_{aj}^{\,\prime\prime(+)}\left(k\right)\vert f^j_{\mathbf{k}\lambda}\left(\mathbf{x}\right)\vert^2 \Big( 2\langle n_{\mathbf{k}\lambda}\rangle + 1\Big)\,,
\\
%%%%%%%%%%%%%%%%%%%%%%%%%%%%%%%%%%%%%%%%%%%%%%%%%%%%%%%%%%%%%%%%%%%%%%%%%
%
\label{a'+-}
\alpha_{aj}^{\,\prime(\mp)}\left(k\right) &=& \sum_b{\alpha_{ab}^{j} k_{ba}\over 2}\left[ \mP {1\over k + k_{ba}}\pm\mP {1\over k-k_{ba}}\right]\,,
\\
%%%%%%%%%%%%%%%%%%%%%%%%%%%%%%%%%%%%%%%%%%%%%%%%%%%%%%%%%%%%%%%%%%%%%%%%%%%%%%%%
%
\label{a''+-}
\alpha_{aj}^{\,\prime\prime(\mp)}\left(k\right) &=& \pi\sum_b{\alpha_{ab}^{j} k_{ba}\over 2}\Big[ \delta\left( k - k_{ba}\right)\pm\delta\left( k + k_{ba}\right)\Big]\,,
\eeq
where $ k = \omega_k/c$, $k_{ab} = \omega_{ab}/c$ and $\langle n_{\mathbf{k}\lambda}\rangle$ is the average number of photons in the mode $\mathbf{k}\lambda$.

Equations (\ref{dEa_rr}-\ref{Qa_fr}) are nothing more that equations (\ref{fr}-\ref{drr}) applied to a neutral but polarizable system interacting with the electromagnetic field. The validity of these equations, once assumed the dipole approximation, is determined by the validity of condition (\ref{Markov}). This implies that the field correlation function in time, $$C^R_{jj}\left(\tau\right) = 2\sum_{\mathbf{k}\lambda}\vert f^j_{\mathbf{k}\lambda}\left(\mathbf{x}\right)\vert^2 \left(\langle n_{\mathbf{k}\lambda}\rangle+{1\over 2}\right)\cos\left(\omega_k\tau\right)\,,$$ obtained from equation (\ref{Ctau}), must have all its characteristic frequencies much larger than any characteristic frequency of the system, so that the correlation time $\tau_c$ obeys the condition: $\tau_c\ll 1/\omega_0$, where $\omega_0$ is the largest characteristic frequency of the system. %If there is any finite characteristic distance $d$, equations (\ref{dEa_rr}-\ref{Qa_fr}) remain valid because this will represent only a constant shift in time, $\tau_c-d/c = \tau_c^{\,\prime}\ll 1/\omega_0$.

Equations (\ref{dEa_rr}) and (\ref{Qa_rr}), give the $(rr)$ contributions to level shifts and energy rates of the system and equations (\ref{dEa_fr}) and (\ref{Qa_fr}) give the $(fr)$ contributions. We see that $(rr)$ contributions do not depend on $\langle n_{\mathbf{k}\lambda}\rangle$, which carries all information about the field state. This information is concentrated on $(fr)$ contributions. Of course, for $\langle n_{\mathbf{k}\lambda}\rangle=0$, there is still a residual term which can be associated with the  contribution of the vacuum fluctuations of the field.

Functions $\alpha_{aj}^{\,\prime(+)}\left(k\right)$ and $\alpha_{aj}^{\,\prime\prime(+)}\left(k\right)$ appearing in the $(fr)$ terms are nothing more that the dispersive and dissipative parts of the susceptibility of the system. In the $(rr)$ contributions, however, we have the functions $\alpha_{aj}^{\,\prime(-)}\left(k\right)$ and $\alpha_{aj}^{\,\prime\prime(-)}\left(k\right)$, which do not have a direct interpretation like their corresponding ``plus" functions. The difference is only in the sign of the $k-k_{ba}$ terms for the first and $k+k_{ba}$ for the second. But this fact will be crucial to determine the behavior of the dispersive interactions: it is precisely this difference of sign which makes possible the existence of the dispersive van der Waals interacion. This fact will become clear when we discuss the two-level system.
\section{Calculation of the dispersive forces}

In this section we shall apply the expressions deduced in the previous section for level shifts and energy rates of a polarizable system interacting with the electromagnetic field. We shall discuss a two-level atom, considered as an isotropic system, in the presence of a perfectly conducting wall.

Let us consider two parallel square plates with sides $L$ fixed at $z=0$ and at $z=\ell$, where $L\gg\ell$, and a small polarizable system on the $\Og Z$ axes at position $z$ that satisfies the condition $0<z\ll\ell$.

Hence, for a given wave-vector $\mathbf{k}$, the quantized electric field (without sources) in the region $0<z<\ell$ is given by \cite{Barton87}:
\beq
\label{ECP}
{\bf E}_{{\bf k}_{\parallel} n}\left( {\bf r}_{\parallel}, z, t \right)&=& \imath\left( \frac{2\pi\hbar k c}{\ell L^2}\right)^{1/2} \Bigg\lbrace a^{(1)}_{{\bf k}_{\parallel}n}\left( {\mathbf{k}_{\parallel}\over k_{\parallel}}\times \mathbf{\hat{z}}\right) \sin \left(\frac{n\pi }{\ell}z\right)\; +
\cr
%%%%%%%%%%%%%%%%%%%%%%%%%%%%%%%%%%%%%%%%%%%%%%%%%%%%%%%%%%%%%%%%%%%%%%%
%
&+&a^{(2)}_{{\bf k}_{\parallel}n}\left[ \imath\frac{n\pi}{k \ell}{\mathbf{k}_{\parallel}\over k_{\parallel}} \sin \left(\frac{n\pi}{\ell}z\right) - \mathbf{\hat{z}} \frac{k_{\parallel}}{k} \cos \left(\frac{n\pi}{\ell}z\right) \right]\Bigg\rbrace  \E{\imath\left( {\bf k}_{\parallel}\cdot {\bf r}_{\parallel} -\omega_k t\right)}\! + \hc \;,\;\;\;\;\;\;\;\;\;\;\;
\eeq
where
\beq
\big[a^{\lambda}_{{\bf k}_{\parallel}n},a^{\lambda^{\prime}}_{{\bf k}^{\prime}_{\parallel}n^{\prime}}\big] &=& \big[a^{\dag {\lambda}}_{{\bf k}_{\parallel}n},a^{\dag {\lambda^{\prime}}}_{{\bf k}^{\prime}_{\parallel}n^{\prime}}\big] = 0\;,
\\
%%%%%%%%%%%%%%%%%%%%%%%%%%%%%%%%%%%%%%%%%%%%%%%%%%%%%%%%%%%%%%%%%%%%%%%%%%%
%
\big[a^{\lambda}_{{\bf k}_{\parallel}n},a^{\dag {\lambda^{\prime}}}_{{\bf k}^{\prime}_{\parallel}n^{\prime}}\big] &=&\delta_{\lambda\lambda^{\prime}}\delta_{n n^{\prime}}\delta_{{\bf k}_{\parallel}{\bf k}^{\prime}_{\parallel}}\;,
\\
%%%%%%%%%%%%%%%%%%%%%%%%%%%%%%%%%%%%%%%%%%%%%%%%%%%%%%%%%%%%%%%%%%%%%%%%%%%
%
{\omega_k^2\over c^2}\;\;=\;\;k^2 &=& k_{_{||}}^2 + \left( n\pi/\ell\right)^2\;,
\\
%%%%%%%%%%%%%%%%%%%%%%%%%%%%%%%%%%%%%%%%%%%%%%%%%%%%%%%%%%%%%%%%%%%%%%%%%%%
%
{\bf k}_{\parallel} = k_x \hat x + k_y \hat y&,&\;\;\;\;\;\;\;\; {\bf r}_{\parallel} = x\, \hat x+ y \,\hat y\;,
\eeq
with $n$ being a non-negative integer number. The mode with $n = 0$ has an additional factor $1/\sqrt{2}$ not shown. Using this expression for the field in the limit $\ell\rightarrow\infty$, we shall be able to calculate the desired dispersive potentials between the atom and the single wall at $z=0$. 

\subsection{Two-level system}

First of all, let us assume isotropy, such that the polarizability given by equation (\ref{a_static}) is independent of the space direction, that is,%
\begin{equation}
\label{a_tl}
\alpha^{j}_{ge} = \alpha_{ge} = -\alpha_{eg} =  {2e^2\over 3\hbar\omega_0}\vert \langle g\vert \mathbf{r}\vert e\rangle\vert^2 = \alpha_0\,,
\end{equation}
where $\vert g\rangle$ and $\vert e\rangle$ are the ground and excited states of the two-level system under consideration with energies $E_g$ and $E_e$, and $\omega_0$ is the transition frequency, given by $\omega_0 = k_0c = \left(E_e-E_g\right)/\hbar$.

Hence, equations (\ref{dEa_rr}) and (\ref{dEa_fr}) for level shifts take the form
\beq
\label{2Level_rr}
\delta E_{g}^{rr}&=&\delta E_{e}^{rr}\; = \;-{1\over 2}\sum_{{\bf k}\lambda}\alpha_{-}^{\,\prime}(k)\vert {\bf f}_{{\bf k}\lambda}\left({\bf x}\right)\vert^2\;,
\\
%%%%%%%%%%%%%%%%%%%%%%%%%%%%%%%%%%%%%%%%%%%%%%%%%%%%%%%%%%%%%%%%%%%%%%%%%%%%%%%%%%%%%%%%%%%
%
\label{2Level_fr}
\delta E_{g}^{fr}&=&-\delta E_{e}^{fr} \;=\;-\sum_{{\bf k}\lambda}\alpha_{+}^{\,\prime}(k)\vert {\bf f}_{{\bf k}\lambda}\left({\bf x}\right)\vert^2\left(\langle n_{{\bf k}\lambda}\rangle+{1\over 2}\right)\;,
\\
%%%%%%%%%%%%%%%%%%%%%%%%%%%%%%%%%%%%%%%%%%%%%%%%%%%%%%%%%%%%%%%%%%%%%%%%%%%%%%%
%
\label{alpha}
\alpha_{\mp}^{\,\prime}\left( k\right)&=&\frac{\alpha_0 k_0}{2}\left(  {\mathcal P}\frac{1}{k + k_0} \pm {\mathcal P}\frac{1}{k - k_0}\right)\;,
\eeq

Note that, while $(rr)$ contribution for level shifts of $\vert g\rangle$ and $\vert e\rangle$ states are exactly the same, $(fr)$ contribution  have the same magnitude but opposite signs. 

Using the expression (\ref{ECP}) in equations (\ref{2Level_rr}-\ref{2Level_fr}), taking the limit $\ell\rightarrow\infty$ and considering a continuous spectra for the field, we are faced with integrals of the form
\beq
\label{A+-}
{\alpha_0 k_0\over 2}\mathcal{A}^{(\pm)}_{\lambda}\left(k_0, f\right) = \int_{0}^{\infty}\dev k\,f\left(k\right) \alpha_{\pm}^{\,\prime}(k)\E{\imath k\lambda}\,,
\eeq
where $\lambda > 0$ is a real parameter. For an analytical function $f$ satisfying the condition
$$%\beq
%
%\label{fanal}
%
\lim_{\vert\im\left[k\right]\vert\rightarrow\infty}\vert f\left(k\right)\vert \E{-\lambda\vert\im\left[k\right]\vert} = 0\,
$$%\eeq
in all complex plane, it is possible to write
\beq
\label{Apm}
\mathcal{A}^{(\pm)}_{\lambda}\left(k_0, f\right) &=& \mp\I\pi f\left(k_0\right)\E{\imath k_0\lambda}+\int_{0}^{\infty}{\dev k\over k+k_0}\left[f\left(k\right)\E{\imath k\lambda}\mp f\left(-k\right)\E{-\imath k\lambda}\right].\;\;\;\;\;\;\;
\eeq
Hence, with the aid of last result, the $z$-dependent part of equations (\ref{2Level_rr}-\ref{2Level_fr}), which gives the dispersive potentials between the atom and the wall, when the average number of photons per mode is independ of the polarization, $\langle n_{\mathbf{k}\lambda}\rangle = \langle n_k\rangle$, takes the form 
\beq
\label{VfCP}
V_g\left(z,\langle n\rangle\right) &=& V_0^{rr}\left(z\right) + V_0^{fr}\left(z\right) + V_{\langle n\rangle}^{fr}\left(z\right)\,,\;\;
\\
%%%%%%%%%%%%%%%%%%%%%%%%%%%%%%%%%%%%%%%%%%%%%%%%%%%%%%%%%%%%%%%%%%%%%%%%
%
\label{VeCP}
V_e\left(z,\langle n\rangle\right)&=&V_0^{rr}\left(z\right) - V_0^{fr}\left(z\right) - V_{\langle n\rangle}^{fr}\left(z\right)\,,\;\;
\\
%%%%%%%%%%%%%%%%%%%%%%%%%%%%%%%%%%%%%%%%%%%%%%%%%%%%%%%%%%%%%%%%%%%%%%%%
%
\label{V0rCP}
V_0^{rr}\left(z\right) &=& {\hbar\omega_0\over 8\pi}{\alpha_0\over z^3}\mathcal H_0^{rr}\left(2k_0z\right)\,,
\\
%%%%%%%%%%%%%%%%%%%%%%%%%%%%%%%%%%%%%%%%%%%%%%%%%%%%%%%%%%%%%%%%%%%%%%%%
%
\label{V0fCP}
V_0^{fr}\left(z\right) &=& {\hbar\omega_0\over 8\pi}{\alpha_0\over z^3}\big[\mathcal H_0\left(2k_0z\right)-\mathcal H_0^{rr}\left(2k_0z\right)\big]\,,
\\
%%%%%%%%%%%%%%%%%%%%%%%%%%%%%%%%%%%%%%%%%%%%%%%%%%%%%%%%%%%%%%%%%%%%%%%%
%
\label{VnfrCP}
V_{\langle n\rangle}^{fr}\left(z\right) &=& {2\hbar c\over \pi}\int_0^{\infty}k^3\alpha_{+}^{\,\prime}\left(k\right)\langle n_k\rangle G\left(2kz\right)\dev k\,,
\\
%%%%%%%%%%%%%%%%%%%%%%%%%%%%%%%%%%%%%%%%%%%%%%%%%%%%%%%%%%%%%%%%%%%%%%
%
\label{Gcp}
G\left(x\right) &=& {\sin x\over x}+2{\cos x\over x^2}-2{\sin x\over x^3}\,,
\\
%%%%%%%%%%%%%%%%%%%%%%%%%%%%%%%%%%%%%%%%%%%%%%%%%%%%%%%%%%%%%%%%%%%%%%%%%%%%
%
\label{H0rr}
\mathcal{H}_0^{rr}\left(x\right)&=&-\pi\left(\cos x+x\sin x -{1\over 2}x^2\cos x\right)\,,
\\
%%%%%%%%%%%%%%%%%%%%%%%%%%%%%%%%%%%%%%%%%%%%%%%%%%%%%%%%%%%%%%%%%%%%%%%%%%%%%%%%%
%
\label{H0}
\mathcal{H}_0\left(x\right) &=& \left(x^2-2\right)\mF\left(x\right)+2x\mG\left(x\right) - x\,,
\\
%%%%%%%%%%%%%%%%%%%%%%%%%%%%%%%%%%%%%%%%%%%%%%%%%%%%%%%%%%%%%%%%%%%%%%%%%%%%%
%
\label{F}
\mF\left(x\right) &=& \Ci\left(x\right)\sin x-\si\left(x\right)\cos x\,,\;\;
%
%\\
%%%%%%%%%%%%%%%%%%%%%%%%%%%%%%%%%%%%%%%%%%%%%%%%%%%%%%%%%%%%%%%%%%%%%%%%
%
%\label{G}
%
\mG\left(x\right) =  {\dev\over\dev x}\mF\left(x\right)\,,%\Ci\left(x\right)\cos x+\si\left(x\right)\sin x\,,
\\
%%%%%%%%%%%%%%%%%%%%%%%%%%%%%%%%%%%%%%%%%%%%%%%%%%%%%%%%%%%%%%%
%
%\label{si}
%
\si\left(x\right) &=& -{\pi\over 2}+\int_0^x \dev t\;{\sin t\over t}\,,\;\;
%
%\\
%%%%%%%%%%%%%%%%%%%%%%%%%%%%%%%%%%%%%%%%%%%%%%%%%%%%%%%%%%%%%%%%%%%%%%%%
%
%\label{Ci}
%
\Ci\left(x\right) = \gamma + \ln x + \int_0^x \dev t\;{\cos t-1\over t}\,,\;\;\;\;\;\;\;\;\;
\eeq
where $V_0^{rr}\left(z\right)$ and $V_0^{fr}\left(z\right)$ are the $(rr)$ and $(fr)$ terms of the vacuum contribution to the interaction and $\gamma$ is the Euler-Mascheronni constant. The term $V_{\langle n\rangle}^{fr}\left(z\right)$ that carries the information about field state (thermal state, for example) and does not contribute when there is no photons in any mode of the field.

Exploring this last case, that is, when $\langle n_k\rangle = 0$, which means that the field is in its vacuum state, we have
\beq
\label{V0f}
V_g\left(z,0\right) &=&  V_0^{rr}\left(z\right)+V_0^{fr}\left(z\right) = V_0\left(z\right)\,,
\\
%%%%%%%%%%%%%%%%%%%%%%%%%%%%%%%%%%%%%%%%%%%%%%%%%%%%%%%%%%%%%%%%%%%%%%%%%%%%%%%%%%%%%%%%%%%%%
%
\label{V0e}
V_e\left(z,0\right) &=& V_0^{rr}\left(z\right)-V_0^{fr}\left(z\right) = 2V_0^{rr}\left(z\right)-V_0\left(z\right)\,,
\eeq
for the ground state and excited state vacuum contributions to dispersive potentials, respectively. Analysing the small and large distance limits, one may show that
\begin{itemize}
\item For $k_0 z\ll 1$ (short distance limit):
\begin{equation}
\label{LVDW}
V_e\left(z,0\right) \simeq V_g\left(z,0\right) = -{\hbar\omega_0\over 8}{\alpha_0\over z^3}+\Og\left(z^{-2}\right)\,.
\end{equation}
\item For $k_0 z \gg 1$ (large distance limit):
\beq
\label{CP}
V_g\left(z,0\right) &=& -{3\hbar c\over 8\pi}{\alpha_0\over z^4}+\Og\left(z^{-6}\right)\,,
\\
%%%%%%%%%%%%%%%%%%%%%%%%%%%%%%%%%%%%%%%%%%%%%%%%%%%%%%%%%%%%%%%%%%%%%%%%%%%%%%%%%%%%%%%%%%%%
%
\label{RL}
V_e\left(z,0\right) &\simeq& {3\hbar c\over 8\pi}{\alpha_0\over z^4}+\hbar c\alpha_0 k_0^4{\cos\left(2k_0 z\right)\over 2k_0 z}\,.
\eeq
\end{itemize}

Equation (\ref{LVDW}) can be recognized as the  interaction between the atom and the wall in the London-van der Waals limit \cite{Len}. In this limit, both excited and ground state potentials coincides and are mainly given by the $(rr)$ contribution (only radiation reaction is important) as one can easily verify from equation (\ref{V0rCP}) when $k_0 z\ll 1$. %This interaction may be interpreted as that between the atom's induced dipole with its own instantaneous image.

The equation (\ref{CP}), which gives the ground state potential at large distance, is the well known Casimir-Polder potential \cite{CasPol1948}. However, by equation (\ref{RL}), we see that in this limit the excited state potential does not coincides with the ground state potential anymore. Apart from a Casimir-Polder term with opposite sign, there is an oscillating one that falls as $1/z$ \cite{Hinds,Barton87} and since the Casimir-Polder term falls as $1/z^4$, the oscillatory term dominates over all large distance limit and the interaction presents an infite number of {\it potential wells}. %

In order to get a better understanding about last results, let us come back to equations (\ref{2Level_rr}-\ref{2Level_fr}). Making $\langle n_{\mathbf{k}\lambda}\rangle = 0$, the ground and excited state level shifts of the atom take the form
\beq
\label{dEnress}
\delta E_g &=& \delta E_g^{fr}+\delta E_g^{rr}\; =\; -{\alpha_0 k_0\over 2}\sum_{\mathbf{k}\lambda}{\vert\mathbf{f}_{\mathbf{k}\lambda}\left(\mathbf{x}\right)\vert^2\over k+k_0}\,,
\\
%%%%%%%%%%%%%%%%%%%%%%%%%%%%%%%%%%%%%%%%%%%%%%%%%%%%%%%%%%%%%%%%%%%%%%%%%%%%%%%%%%%%%%%%%%%%%
%
\label{dEress}
\delta E_e &=& \delta E_e^{fr}+\delta E_e^{rr}\; =\; -{\alpha_0 k_0\over 2}\mP\sum_{\mathbf{k}\lambda}{\vert\mathbf{f}_{\mathbf{k}\lambda}\left(\mathbf{x}\right)\vert^2\over k-k_0}\;.
\eeq

From last equations we clearly see that the potential associated to the ground state depends only on the {\it non-resonant} part of the atomic polarizability, while the potential associated to the excited state depends only on the {\it resonant} part. Since van der Waals forces are due to level shifts of the ground states of the interacting atoms \cite{Holstein} it may be called {\it dispersive non-resonant interaction}. On the other hand, for obvious reasons, the excited state potential may be called as {\it dispersive resonant interaction}.

Last equations permit us also to understand why resonant interaction is very much stronger than van der Waals interaction at large distances and practically equal to this last at short distances. The  position dependent part of the modulus square of the amplitute of a field mode (subject to boundary conditions) which appears in the sums of the equations (\ref{dEnress}-\ref{dEress}), is usualy an oscillating function of products of characteristic distances and the frequency associated to the mode. In our case the characte\-ris\-tic distance is $z$, which means that for high frequencies such that $k > 2\pi/z$, the oscillatory behavior of the summands leads to a small contribution to the total sum compared to the contribution comming from the low frequencies ($k < 2\pi/z$). 

Taking into account that the number of modes per frequency is proportional to $k^2$, at short distances virtual photons with high frequencies $(k\sim 2\pi/z\gg k_0)$ are much more important to the interaction than those with low frequencies $(k \ll k_0)$, so that $k_0$ may be neglected in denominators of the sums in equations (\ref{dEnress}-\ref{dEress}) and the level shifts will be approximately the same. However, as the distance between the atom and wall becomes large, smaller and smaller frequencies become important to the interaction. For the ground state contribution, this leads to a smooth variation of the interaction. However, for the excited state contribution, when the distance becomes of the order $2\pi/k_0$, there is a strong increasing of the interaction, so the summand becomes singular at the frequency $\omega_0$. At very large distances, small frequencies become more important and both ground and excited state potentials become weak, but the excited state potencial still remains very much stronger than the ground state potential. This happens because though the oscillations in summands tend to cancel contributions of frequencies $ 2\pi/z<k\ll k_0$, the density number of modes is too small at very low frequencies, which still makes the principal value in (\ref{dEress}) the main contribution.

One can get a simple interpretation of this fact. Looking at equation (\ref{RL}) more carefully we  easily see that a maximum and a minimum of the potential occur, respectively, at

\beq
\label{mxmn}
z \simeq n\,{\lambda_0\over 2}\;\;\;(\mathrm{maximum})\;,\;\;\;\;\;\;
z \simeq \left(n-{1\over 2}\right){\lambda_0\over 2}\;\;\;(\mathrm{minimum})\,,
\eeq
where $\lambda_0=2\pi c/\omega_0$ is the transition wavelength and $n\geq 1$ is an integer (of course, these approximations will be better for higher values of $n$). Equations (\ref{mxmn}) are nothing more than the conditions for the existance of {\it stationary} waves with wavelength $\lambda_0$ in the ``cavity" defined by the atom and the wall: the former is the resonant condition for a one-dimensional cavity with the ``closed ends" and the latter is the resonant condition for a cavity with one ``closed end" and one ``opened end". Then, it is comprehensible why the interacting potentials have the maximum magnitude or, in other words, a maximum response at that distances and not at others.
\begin{figure}[!h]
\begin{center}
\includegraphics[width=4.0in]{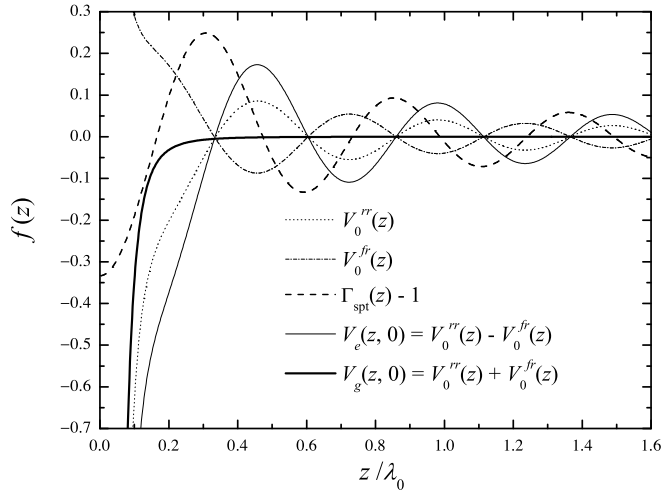}
\caption{Dispersive potentials associated to the ground (van der Waals interaction) and excited (resonant interaction) states with corresponding $(fr)$ and $(rr)$ contributions. The total spontaneous emission rate as function of $z$, $\Gamma_{\mathrm{spt}}\left(z\right)$, is also shown. The potentials are in units of $\hbar c\alpha_0 k_0^4$ and the spontaneous rate is in units of $2c\alpha_0 k_0^4$, which is the value for the spontaneous rate in the free space.}
\label{V0}
\end{center}
\end{figure}

In Figure \ref{V0} we show the exact potentials given by equations (\ref{V0f}-\ref{V0e}) and the $(rr)$ and $(fr)$ contributions given by equations (\ref{V0rCP}-\ref{V0fCP}). As we pointed out previously, the ground and excited state potentials are approximately the same as the London-van der Waals potential (\ref{LVDW}) at short distances. At large distances, the excited state potencial has an oscillatory behavior while the ground state potential is given by the Casimir-Polder potential (\ref{CP}). 

We saw that at short distances the radiation-reaction on the polarizable system is the dominant contribution to the interaction, a well established result in literature \cite{Morawitz,Meschede,Hinds}. However, Casimir-Polder potential is usually interpreted as a stark shift due to the field fluctuations mainly \cite{Hinds}. It is a common thing the use of expressions similar to equation (\ref{2Level_fr}),  for $\langle n_{\mathbf{k}\lambda}\rangle = 0$, in order to calculate the Casimir-Polder potential only replacing the atomic polarizability $\alpha_{+}^{\,\prime}\left(k\right)$ by its static value $\alpha_{+}^{\,\prime}\left(0\right) = \alpha_0$ in the summation \cite{MilonniLivro}. This procedure can be justified if we invoke the argument that frequencies smaller than $k\sim 2\pi /z\ll k_0$ give the main contribution to the interaction, so that the substitution of $\alpha_{+}^{\,\prime}\left(k\right)$ by $\alpha_0$ causes a little error but gives the correct leading term of the potential. 

Now, despite the success of this procedure, this seems to hide the true physical content behind it. As one can see from Figure \ref{V0}, both $(rr)$ and $(fr)$ contributions are of the same order of magnitude for distances $z > 0.2\lambda_0$, but have opposite signs. For this reason, the excited state interaction given by the equation (\ref{V0e}) is about two times ($rr$) contribution in the  above mentioned interval. However, for the ground state potential given by the equation (\ref{V0f}), the $(rr)$ and $(fr)$ contributions almost cancel to each other, leading to a small potential value which is precisely the van der Waals interaction, which reduces to Casimir-Polder potential for $z>\lambda_0$. Then, {\it both} vacuum field fluctuations and radiation-reaction play a crucial role for the interaction at large distances and the interpretation of the Casimir-Polder interaction as a shift due only to field fluctuations is not complete.

%
%
%\subsubsection{Exchange energy rates}

In order to conclude this section, let us make a comment about the exchange energy rates for the two-level system. 

From equations (\ref{Qa_rr}) and (\ref{Qa_fr}), one may easily show that $\Q_g = 0$ and
\beq
\Q_e = -\pi\alpha_0 k_0^2 c\sum_{\mathbf{k}\lambda}\vert\mathbf{f}_{\mathbf{k}\lambda}\left(\mathbf{x}\right)\vert^2\delta\left(k-k_0\right)\,,
\eeq
where we considered the field in its vacuum state. For the geometry treated here, namely, with two-level atom at a distance $z$ from an infinite perfectly conducting wall, last equation leads to \cite{MilonniLivro}
\beq
\label{QeTL}
\Q_e = -\hbar\omega_0\Gamma_{\mathrm{spt}}\left(z\right)\;,\;\;\;\Gamma_{\mathrm{spt}}\left(z\right)=\Gamma_{e\rightarrow g}^{\mathrm{spt}}\big[1-G\left(2 k_0 z\right)\big]\,,
\eeq
where $\Gamma_{\mathrm{spt}}\left(z\right)$ is the total spontaneous emission rate of the atom in the presence of the wall and $$\Gamma_{e\rightarrow g}^{\mathrm{spt}} = 2 c\alpha_0 k_0^4 = {4\alpha\over 3 c^2}\vert\langle e\vert\mathbf{r}\vert g\rangle\vert^2\omega_0^3$$
is the spontaneous emission for the atom in the empty space, being $\alpha = e^2/\hbar c\simeq 1/137$ the fine structure constant. Then, the life time of the excited state of the atom is of the order $1/\Gamma_{\mathrm{spt}}\left(z\right)$, so that the resonant interaction is unstable and at a finite time (generally about $10^{-7}\mathrm{s}$ in the visible region \cite{Nist}) the van der Waals interaction takes place.

Equation (\ref{QeTL}) for $\Gamma_{\mathrm{spt}}\left(z\right)$ is also plotted in the Figure \ref{V0}. The behavior of the spontaneous emission is similar to the behavior of the resonant potential. However, we should note that there is a difference of phase between the potential and the spontaneous emission of approximately $\pi/2$. Correspondimg to a maximum or a minimum of the potential, the spontaneous emission is exactly the same as if the atom was in the empty space. Somehow this could have been anticipated once for these cases, from conditions (\ref{mxmn}), there is a node on the wall. 

A final comment is in order here: the fact that the potential is in quadrature with the spontaneous emisson, may be understood if we remember that these quantities are related to real and imaginary parts, respectively, of the susceptibilities (of the atom and the field) and these quantities are, of course, in quadrature.

\subsection{Thermal corrections to the interaction}

Let us now turn our attention to the thermal corrections to dispersive interactions by considering both ground and excited states of the atom. There are many works in the literature treating thermal corrections to dispersive van der Waals forces, like \cite{Lifshitz, FeinSucher,BartCPFS} to mention just a few. However, we have not found works that have analyzed also thermal corrections to the resonant interaction or the high temperature limit for dispersive interaction between the two level system and the perfectly conducting wall. This is our aim in this section. 

From equations (\ref{VfCP}-\ref{VeCP}), one may see that the only term that accounts for corrections to the interaction due to non-trivial states of the field is $V^{fr}_{\langle n\rangle}\left(z\right)$, which is given by equation (\ref{VnfrCP}). For a thermal state of the field at temperature $T$, we write
\beq
\label{VTz}
V^{fr}_{\langle n\rangle}\left(z\right) = V_T\left(z\right) = \frac{2\hbar c}{\pi}\int_0^\infty \frac{k^3 \alpha_{+}^{\,\prime}\left( k\right)}{ \E{k\lambda_T} - 1}\,G\left( 2 k z\right)\dev k\,,
\eeq
where $\lambda_T = \hbar c/k_B T$ is the {\it thermal length}, which defines the length scale beyond which thermal contributions pass to dominate over vacuum contribution of the van der Waals interaction. %\cite{Mostepa}. %
At room temperature, $\lambda_T\simeq 7.63\mu\mathrm{m}$.

In the low temperature limit, $k_0\lambda_T\gg 1$, we have already caculated the last integral in both small and large distance limits in a previous work \cite{TarFarJPA2006}, which for thermal corrections are defined by conditions $z\ll\lambda_T$ and $z > \lambda_T$ respectively.

For $z\ll\lambda_T$, we have
\beq
\label{VTzsmall}
V_T\left(z\right)\simeq C\left(T\right) - {\left(2\pi\right)^5\over 315}{\hbar c\alpha_0\over\lambda_T^6}z^2\,,
\eeq
where $C(T) = 2\pi^3\hbar c\alpha_0/45\lambda_T^4$ is independent of $z$ and does not contribute to the force. The above result, that represents a small correction for the vacuum contribution, fits well the exact potential given by (\ref{VTz}) within an error smaller than $3\%$ for $z < 0.10\lambda_T$ and, since it does not depend  explicitly on $\lambda_0$, it is valid for both London-van der Waals and Casimir-Polder limits%
\footnote{For the Casimir-Polder limit, the condition $\lambda_0<z\ll\lambda_T$ must be valid.}.%

For $z\sim\lambda_T$ or larger, we have 
\beq
\label{LifAsynResult}
V_0\left(z\right)+V_T\left(z\right) \simeq V_{\mathrm{Lif}}\left(z,T\right)= -{k_B T\over 4}{\alpha_0\over z^3}\, ,
\eeq
which is exactly the Lifshitz asymptotic result \cite{Lifshitz}, which we hereafter call only {\it Lifshitz's potential} for simplicity. If we consider only the ground state interaction, last equation fits the exact potential within an error smaller than $0.1\%$ for $z>\lambda_T$. 

The ground and excited state potentials with the corresponding thermal contributions are
\beq
\label{VgT}
V_g\left(z,T\right) &=& V_0^{rr}\left(z\right)+V_0^{fr}\left(z\right)+V_T\left(z\right)\; =:\; V\left(z,T\right)\,,
\\
%%%%%%%%%%%%%%%%%%%%%%%%%%%%%%%%%%%%%%%%%%%%%%%%%%%%%%%%%%%%%%%%%%%%%%%%%%%%%%%%%%%%%%%
%
\label{VeT}
V_e\left(z,T\right) &=& V_0^{rr}\left(z\right)-V_0^{fr}\left(z\right)-V_T\left(z\right)\; =\; 2V_0^{rr}\left(z\right)-V\left(z,T\right)\,,
\eeq
where $V(z,T)$ is defined by equation (\ref{VgT}). Considering the atom in thermal equilibrium with the field, we must consider the thermal average value of last potentials, since this average is the  observed potential. If $p$ is the probability of the atom to be found in its ground state, $1-p$ is the probability of finding it in its excited state. In a thermal equilibrium at a temperature $T$, we have
$$ p = {1\over 1+\E{-k_0\lambda_T}}\,,$$
so that the average potential takes the form
\beq
\label{VmeanTot}
\bar V\left(z,T\right) = \tanh\left({1\over 2}k_0\lambda_T\right)V\left(z,T\right) + {2V_0^{rr}\left(z\right)\over \E{k_0\lambda_T}+1}\,.
\eeq

At low temperature, $k_0\lambda_T\gg 1$, one has
$$\bar V_{\mathrm{low}}\left(z,T\right)\simeq V\left(z,T\right)-2\E{-k_0\lambda_T}\left[V_0^{fr}\left(z\right)+V_T\left(z\right)\right]\,.$$
The average potential is given by the ground state potential minus the $(fr)$ contribution weighted by the exponential factor $\E{-k_0\lambda_T}\ll 1$, so that the average potential is practically due to the ground state potential only, as expected: the thermal photons do not have enough energy to excite the transition and populate the excited state, which makes negligible its contribution to the interaction. For example, at room temperature and for a transition frequency in the visible region, $k_0\simeq 10\mu\mathrm{m}^{-1}$, the factor multiplying the $(fr)$ term is about $7.1\times 10^{-34}$. %Hence, expression (\ref{LifAsynResult}) for large distances $(z > \lambda_T)$ and the expression (\ref{VTzsmall}) summed to potential (\ref{LVDW}), for $k_0 z\ll 1$, or summed to (\ref{CP}), for $k_0 z\gg 1$, when the distance satisfies the condition $z\ll\lambda_T$, describe completely the interaction.

For the high temperature limit, however, things are diffent. In this limit, the excited state gives an important contribution to the interaction. In order to account the excited state influence on the interaction at high temperature, $k_0\lambda_T\ll 1$, it is necessary recalculate the integral in (\ref{VTz}) with the help of Bernoulli's numbers, defined by the expansion \cite{Arfken}
$$ {\xi\over\E{\xi}-1} = 1-{1\over 2}\xi+\sum_{n=1}^{\infty}{B_{2n}\over\left(2n\right)!}\xi^{2n}\,.$$
Using last expression in equation (\ref{VTz}), we obtain
\beq
\label{VTHIGH}
V_T\left(z\right) &=& \hbar\Delta_T^{(1)} + \hbar\Delta_T^{(2)} + \hbar\Delta_T^{(3)}\;,
\\
%%%%%%%%%%%%%%%%%%%%%%%%%%%%%%%%%%%%%%%%%%%%%%%%%%%%%%%%%%%%%%%%%%%%%%%%%%%%%%%
%
\label{hDT(1)}
\hbar\Delta_T^{(1)} &=& {k_B T\over 4\pi z^3}\int_0^{\infty}x^2\alpha_+^{\,\prime}\left(x\right)G\left(x\right)\dev x\;,
\\
%%%%%%%%%%%%%%%%%%%%%%%%%%%%%%%%%%%%%%%%%%%%%%%%%%%%%%%%%%%%%%%%%%%%%%%%
%
\label{hDT(2)}
\hbar\Delta_T^{(2)} &=& -{\hbar c\over 16\pi z^4}\int_0^{\infty}x^3\alpha_+^{\,\prime}\left(x\right)G\left(x\right)\dev x = -V_0^{fr}\left(z\right)\;,
\\
%%%%%%%%%%%%%%%%%%%%%%%%%%%%%%%%%%%%%%%%%%%%%%%%%%%%%%%%%%%%%%%%%%%%%%%%
%
\label{hDT(3)}
\hbar\Delta_T^{(3)} &=& {k_B T\over 4\pi z^3}\sum_{n=1}^{\infty}{B_{2n}\over \left(2n\right)!}\,\eta^{2n}\!\int_0^{\infty}x^{2n+2}\alpha_+^{\,\prime}\left(x\right)G\left(x\right)\dev x\;,
\eeq
where $\eta = \lambda_T/2z$ and $x = 2kz$. Integrals in equation (\ref{hDT(3)}) may be easily calculated with the help of equation (\ref{Apm}), which leads to
\beq
\hbar\Delta_T^{(3)} = V_0^{rr}\left(z\right)\left[{2\over k_0\lambda_T}-\coth\left({1\over 2}k_0\lambda_T\right)\right]\,.
\eeq
\begin{figure}[!h]
\begin{center}
\includegraphics[width=4.0in]{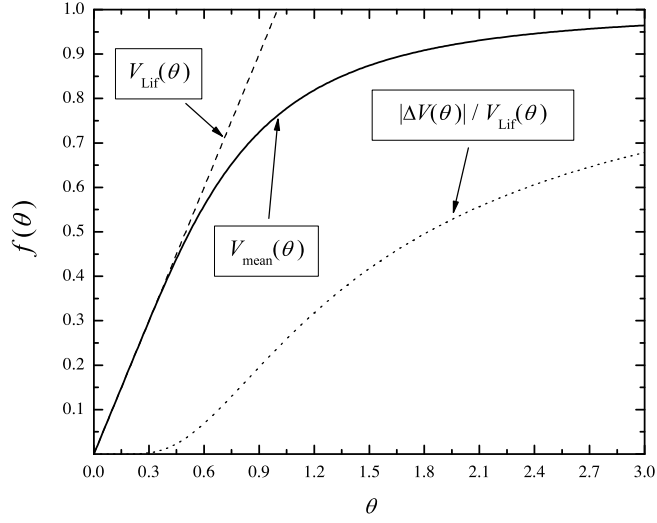}
\caption{Dispersive potentials between a two-level atom and a perfectly conducting wall in a thermal bath at temperature $T$. We plot the temperature dependence of the thermal averaged potential $\bar V\left(z,T\right)$, as well as the potential associated to the ground state only  $V_{\mathrm{Lif}}\left(z,T\right)$, normalized by $-\hbar\omega_0\alpha_0/8z^3$ ($V_{\mathrm{mean}}\left(\theta\right)$ and $V_{\mathrm{Lif}}\left(\theta\right)$ respectively). The relative error $\vert \Delta V\left(\theta\right)\vert/V_{\mathrm{Lif}}\left(\theta\right)$ is also plotted. In the above figure it is assumed that $z > \lambda_T$ and $\theta = 2k_B T/\hbar\omega_0$.}
\label{VmT}
\end{center}
\end{figure}

The integral in equation (\ref{hDT(1)}) may be also calculated in a simple manner from equation (\ref{Apm}), but note that the function $f(x)=x^2 G(x)$ is not analytical in $x=0$ and a direct application of (\ref{Apm}) depends on the analyticity of $f(x)$ over all complex plane. This problem may be avoided by splitting the integral in (\ref{hDT(1)}) as follows
\beq
\int_0^{\infty}x^{2}\alpha_+^{\,\prime}\left(x\right)G\left(x\right)\dev x
%
%\cr
%%%%%%%%%%%%%%%%%%%%%%%%%%%%%%%%%%%%%%%%%%%%%%%%%%%%%%%%%%%%%%%%%%%%%
%
\label{I(1)bottom}
= \pi\alpha_0\left(x_0\sin x_0-{1\over 2}x_0^2\cos x_0\right)-2\int_0^{\infty}\alpha_+^{\,\prime}\left(x\right){\sin x\over x}\,\dev x\,,
\eeq
where $x_0 = 2k_0 z$. Last term in the previous equation can be calculated as
\beq
\label{a+sx/x}
\int_0^{\infty}\alpha_+^{\,\prime}\left(x\right){\sin x\over x}\,\dev x = \alpha_0\int_0^{\infty}{\sin x\over x}\,\dev x-{1\over x_0}\int_0^{\infty}\alpha_-^{\,\prime}\left(x\right)\sin x\,\dev x = {\pi\alpha_0\over 2}\left(1-\cos x_0\right)\;.\;\;
\eeq

Combining equations (\ref{a+sx/x}) and (\ref{I(1)bottom}), equation (\ref{hDT(1)}) may be written as
\beq
\label{DT1Result}
\hbar\Delta_T^{(1)} = -{k_B T\over 4}{\alpha_0\over z^3}-{2\over k_0\lambda_T}V_0^{rr}\left(z\right)\,
\eeq
and the ground state potential (\ref{VgT}) takes the form
\beq
\label{VzTsf}
V\left(z,T\right)=  -{k_B T\over 4}{\alpha_0\over z^3}-{2V_0^{rr}\left(z\right)\over \E{k_0\lambda_T}-1}\,.
\eeq

Inserting last result into equation (\ref{VmeanTot}), the average potential may finally be written as
\beq
\label{Vtheta}
\bar V\left(z,\theta\right) = -{\hbar\omega_0\over 8}{\alpha_0\over z^3}\,\theta\tanh\left(1/\theta\right)\,,
\eeq
%%%%%%%%%%%%%%%%%%%%%%%%%%%%%%%%%%%%%%%%%%%%%%%%%%%%%%%%%%%%%%%%%%%%%%%%%
%
where we defined the normalized temperature $\theta = 2k_B T/\hbar\omega_0 = 2/k_0\lambda_T$.

We derived equation (\ref{Vtheta}) by assuming the high temperature limit, which means  $\theta\sim 1$ or higher. However, if we consider the opposite limit, $\theta\ll 1$, we  easily see that equation (\ref{Vtheta}) reduces to equation (\ref{LifAsynResult}), $$\bar V\left(z,T\right) = -{k_B T\over 4}{\alpha_0\over z^3}\left(1-2\E{-k_0\lambda_T}+2\E{-2k_0\lambda_T}+...\right)\simeq -{k_B T\over 4}{\alpha_0\over z^3}\,,$$ which is valid for all distances larger than the thermal length, that is, for $z > \lambda_T$. Hence, equation (\ref{Vtheta}), which takes into account the excited state contribution, is also valid for all distances larger than the thermal length.
\begin{figure}[!h]
\begin{center}
\includegraphics[width=3.75in]{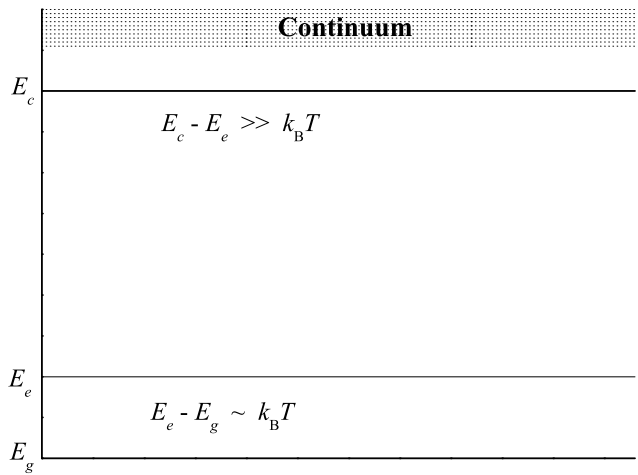}
\caption{A schematic spectrum of a system which simulates a two-level one and that can be used to check our result (\ref{Vtheta}).  Levels $E_e$ and $E_g$ are sufficiently close to each other so that thermal fluctuations can populate significantly both of them: $E_e-E_g \sim k_B T\simeq 1/40\; \textrm{eV}$ at room temperature. Level $E_c$ is so far from $E_e$ and $E_g$  ($E_c-E_e \gg k_B T$) that, in practice, it can not be populated by thermal fluctuations.}
\label{gap}
\end{center}
\end{figure}

In Figure \ref{VmT} we plot the potentials (\ref{Vtheta}) and (\ref{LifAsynResult}) normalized by the  potential (\ref{LVDW}). At low temperatures, as we have already anticipate, the excited state plays no role in the interaction and the ground state potential, which is given by Lifshitz's potential (\ref{LifAsynResult}), coincides with the average potential. However, at high temperatures, the difference between the Lifshitz's and the average potentials is evident, as can be checked in Figure \ref{VmT} for values of $\theta$ greater than $\theta\simeq 0.4$ approximately, which is equivalent to a temperature about $2800 K\times 1\mu\mathrm{m}/\lambda_0$; at this point the relative error between the two potentials, defined as $\vert \Delta V\vert/V_{\mathrm{Lif}}$, is about $1.3\%$ (for $\theta = 1.0$ the error is already $\simeq 24\%$). 

Nevertheless, the most interesting thing that happens as the temperature becomes higher is that the average potential saturate at a value independent of the temperature. In this case, the potential can be approximated by%  %
\beq
\label{VTHigh}
\bar V\left(z,\theta\right)= -{\hbar\omega_0\over 8}{\alpha_0\over z^3}\left(1-{1\over 3\theta^2}+{2\over 15\theta^4}-...\right)\,,
\eeq
which coincides exactly with the potential (\ref{LVDW}) in the limit of very high temperature, $\theta\rightarrow\infty$. 

As mentioned before, expression (\ref{LVDW}) gives the vacuum term of the dispersive potential between the atom and the wall, for both ground and excited states of the atom, in the small distance limit $(0<z\ll\lambda_0)$. Further, when the system interacts with the field in a thermal state, thermal corrections to the potential become important only for distances larger than thermal length $\lambda_T$. Equation (\ref{VTHigh}) is approximately valid for $z>\lambda_T$ at very high temperature, $\theta\gg 1$. This condition necessarily implies $\lambda_T\ll\lambda_0$. Hence, combining last condition with the two just mentioned conditions for the separation $z$ in which pontentials (\ref{LVDW}) and (\ref{VTHigh}) may be applied, one can see that London-van der Waals potential describes the interaction at very high temperature in the interval $0<z<\infty$. In other words, for very high temperatures, the London-van der Waals potential is valid for all distance regimes. This is a quite curious result.

However, though expressions (\ref{LVDW}) and (\ref{VTHigh}) are practically the same, there are some important differences in the physics behind them. The former is interpreted as the interaction between the instantaneous dipole and its mirror image so that retardation does not play any role at all on interaction and, as already mentioned, the main contribution is due to the radiation reaction. 
The second one is the net result of the competition of the fluctuations of the field acting on  the ground and excited states of the atom simultaneously. 

From equations (\ref{VgT}-\ref{VeT}), one can see that thermal corrections have opposite signs and since at high temperature the probability of finding the system in its ground or excited states is practically the same, $p\simeq 1-p\simeq 0.5$, the $(fr)$ terms cancel each other and the temperature dependence tends to disappear in the average potential, as it occurs in (\ref{VTHigh}). We also see that in equations (\ref{VgT}) and (\ref{VeT}) the $(rr)$ terms are the same, which leads to a net average potential different of zero. Hence, both equations (\ref{LVDW}) and (\ref{VTHigh}) come from of the radiation reaction, though from different ways. The fact that retardation does not seem to have any effect on the average potential is a feature of thermal fluctuations which, for distances larger than $\lambda_T$, destroy the influence of the finiteness of the velocity of light on the interaction, as one can see also in (\ref{LifAsynResult}).

As a final comment, one may argue that equation (\ref{VTHigh}) is an unrealistic one because there are not systems in the real world which can be treated as a two-level system at very high temperature. Actually, all known systems usually have an infinite number of energy levels and the coupling between them at high temperature should not be neglected, so that a different behavior of that preconized for the two-level system is expected \cite{Barton74}. Eventhough, the two-level system model showed to be a very useful one in the study of the dispersive interactions and, particularly, in the comprehension of the roles played by $(rr)$ and $(fr)$ contributions. 

Though equation (\ref{VTHigh}) is, in some sense, meaningless, the same can not be said to equation  (\ref{Vtheta}). It is plausible a system where the two lowest levels $\vert g\rangle$ and $\vert e\rangle$ are far enough to a third level $\vert c\rangle$, as sketched in the Figure \ref{gap}. If the largest dimension $D_0$ of this system is much smaller than the transition wavelength, $D_0 \ll 2\pi\hbar c/\left(E_e-E_g\right)$, and the temperature satisfies the condition $E_e-E_g\sim k_B T\ll E_c-E_e$, then the equation (\ref{Vtheta}) will be valid and a deviation of the behavior predicted by equation (\ref{LifAsynResult}) should be observed for not too high values of temperature.

\section{Conclusions}

In this work we made use of the general expressions for level shifts obtained from {\it master equation} in order to study the dispersive potentials between a two-level atom and a perfectly conducting wall in the dipole approximation. We studied the potentials for the ground and excited states of the atom, which can be associated to the van der Waals and the re\-so\-nant interactions, respectively. %
All distance regimes as well as the low and high temperature limits were treated. Considering the ground state potentials in the short distance limit, we reobtained the London-van der Waals potential given by (\ref{LVDW}), showing that it may be explained mainly by the reservoir reaction  (radiation reaction) contribution to the level shift, as it is well accept \cite{Hinds}. %
For large distances, we rederived the Casimir-Polder interaction and showed that it can not be considered as a direct result of the vacuum fluctuations  only, as usually intrepreted \cite{MilonniLivro}. Otherwise, we showed that the field fluctuations and the radiation reaction   contributions are of the same order and too much stronger than Casimir-Polder potential, but these effects ``cancels each other" leaving a small observed interaction. %
For the case of the excited state potential, we emphasized that the maximum magnitudes of the potential occur approximately at the positions of resonant condition for stationary waves with wavelength $\lambda_0$ in the ``cavity" defined by the atom and the wall. %
Taking into account the thermal corrections to the interaction, we rederived the short and large distance potentials at low temperatures, where the excited state contribution can be disregarded. However, for high temperatures, since the excited state contribution should not be neglected anymore, we showed that its inclusion may cause considerable deviations from Lifshitz asymptotic potential. A curious result that we found is that at very high temperatures the potential saturate in a temperature independent value exactly equal to the London-van der Waals potential for all distance regimes. 
\vskip 0.5 cm
\noindent
{\large {\bf Acknowledgment}}
\vskip 0.5 cm
\noindent
The authors are grateful to Faperj and CNPq for the financial support.
\footnotesize{
{%

}
}

\end{document}